\documentclass[lettersize,journal]{IEEEtran}
\usepackage{amsmath,amsfonts}
\usepackage{algorithm}
\usepackage{algpseudocode}
\usepackage{array}
\usepackage{textcomp}
\usepackage{stfloats}
\usepackage{url}
\usepackage{verbatim}
\usepackage{graphicx}
\usepackage{cite}
\usepackage{booktabs}
\usepackage{tikz}
\usepackage{pgf-pie} 
\usepackage{subcaption}
\usepackage{amsthm}
\usepackage{mathtools}
\usepackage{pgfplots}
\usepackage{censor}
\theoremstyle{definition}

\DeclarePairedDelimiter\floor{\lfloor}{\rfloor}

\pgfplotsset{compat=1.17}
\begin{document}

\title{DSP-Packing: Squeezing Low-precision Arithmetic into FPGA DSP Blocks}

\author{Jan~Sommer, Akif~\"Ozkan, Oliver~Keszocze,~\IEEEmembership{Member,~IEEE} J\"urgen~Teich,~\IEEEmembership{Fellow,~IEEE}
\thanks{The authors are with the Friedrich-Alexander-Universit\"at
Erlangen-N\"urnberg, 91054 Erlangen, Germany. E-mail: \{jan.sommer, oliver.kescoeze, akif.oezkan\}@fau.de, teich@cs.fau.de.}}



\maketitle
\begin{abstract}
The number of Digital Signal Processor (DSP) resources available in Field Programmable Gate Arrays (FPGAs) is often quite limited. Therefore, full utilization of available DSP resources for the computationally intensive parts of an algorithm is paramount for optimizing the non-functional properties of an implementation (i.e., performance, power, and area). The DSPs available in Xilinx devices implement large bit width operators (i.e. a 48-bit accumulator or a $\mathbf{18 \times 27}$ multiplier). However, using such a DSP for low-precision quantized data (as is common in image processing or machine learning applications) leaves the DSP resources underutilized. As a remedy, A method has been proposed to pack and compute four 4-bit multiplications on a single DSP in a single clock cycle. This paper presents a generalization of this scheme to arbitrary bit widths and number of multiplications. We also demonstrate that the previously proposed approach leads to errors (Mean Absolute Error (MAE) = 0.37). Furthermore, we explain where these errors come from and how they can be corrected. On top, we introduce a novel approximate method called ``Overpacking'' which allows to squeeze even more multiplications into a single DSP at the cost of small errors (MAE = 0.47). Overpacking allows to squeeze six 4-bit multiplications into a single DSP compared to just four in the literature. Finally, we introduce an alternative method for packing multiple small-bit width additions into a single 48-bit accumulator for use in applications such as Spiking Neural Networks. 
\end{abstract}


\section{Introduction}
\begin{figure}[!t]
  \centering
  \includegraphics[width=0.70\linewidth]{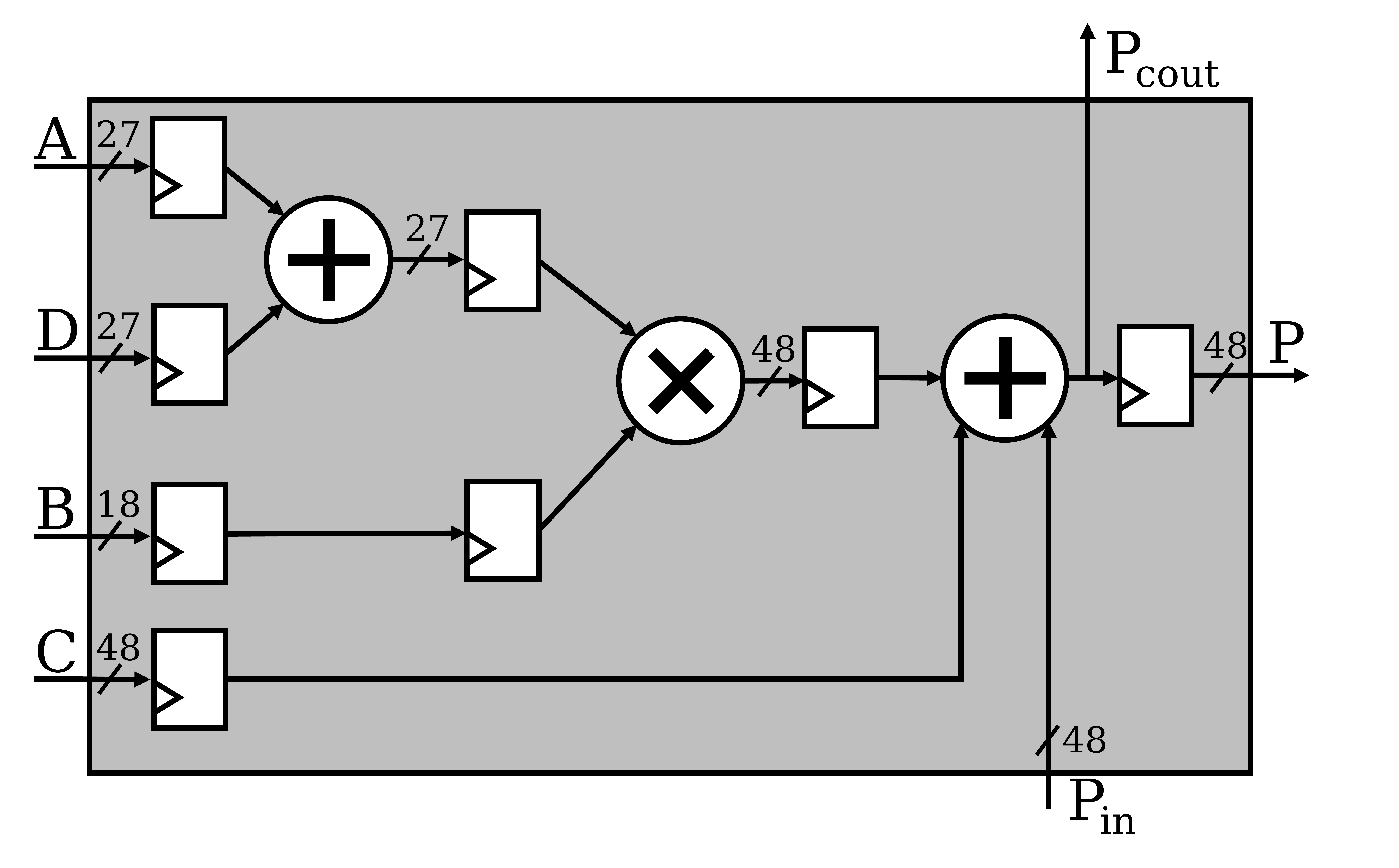}
  \caption{Schematic architecture and dataflow of the Xilinx DSP48E2.}
  \label{fig:DSP48E2}
\end{figure}

Modern FPGAs have a heterogeneous architecture. They consist of programmable fabric, i.e., Lookup Tables (LUTs) and Flip-Flops (FFs), as well as non-programmable hard blocks like dedicated Digital Signal Processors (DSPs). The micro-architecture of such a DSP-block is vendor-dependent. The DSP48E2 that is present in Xilinx UltraScale FPGAs features a 27 bit preadder, a $18 \times 27$ bit multiplier and a 48 bit accumulator~\cite{Xilinx.DSP48E2}. A single DSP48E2 block can implement functions of the form
\begin{equation}\label{eqn:DSP48E2}
    P = B \times (A + D) + C + P_{in}
\end{equation}
as shown in Fig.~\ref{fig:DSP48E2}.

The DSP hard blocks can be used to implement arithmetic circuits that achieve a much better performance (in terms of speed, area and energy efficiency) compared to implementations that use the standard programmable FPGA fabric. Therefore, when designing arithmetic circuits for optimal results, maximal usage of the available DSP blocks is important~\cite{Xilinx.UG479}.

\IEEEpubidadjcol 

Since the DSP blocks are a scarce resource on an FPGA chip, it is important to use them as efficiently as possible. This is a problem for small bit width arithmetic, because implementing such a function on a DSP block would leave most of the DSP's resources unused. Applications in the domain of image processing or machine learning typically operate on quantized data with small bit widths of 8 bit or less~\cite{Xilinx.INT8, Xilinx.INT4}. To improve the utilization of the DSP's resources for low-precision arithmetic, some techniques have been proposed to map multiple small bit width multiplications to a single DSP~\cite{Xilinx.INT8, Xilinx.INT4}. An even better utilization can be achieved when applying Approximate Computing principles. Approximate Computing is an emerging paradigm where non-functional requirements like chip area, latency or energy efficiency are traded of for computational accuracy~\cite{Han.2013, Mittal.2016}. 
Our contributions are as follows:
\begin{itemize}
    \item We generalize the multiplication packing technique proposed by Xilinx Inc.~\cite{Xilinx.INT8, Xilinx.INT4} to arbitrary bit widths and arbitrary number of multiplications independent of the DSP's architecture.
    \item We show how multiplication packing technique proposed by Xilinx Inc.~\cite{Xilinx.INT8, Xilinx.INT4} leads to small errors and how these errors can be corrected.
    \item We present a novel packing strategy called ``Overpacking'' to squeeze more (logical) values into one (physical) DSP input such that partially incorrect results are produced. We show techniques how the erroneous results can be improved using additional logic. Overpacking allows up to 50\% more multiplications on a single DSP compared to the approaches proposed by Xilinx Inc.~\cite{Xilinx.INT8, Xilinx.INT4}.
    \item We also present a strategy for packing small bit width additions into the DSPs 48-bit accumulator.
\end{itemize}

\section{Related Work}
Huang et al.~\cite{Huang.2019} propose a method for implementing parallel multiplications on a single DSP slice. On the DSP48E2 that is present on Xilinx UltraScale FPGAs~\cite{Xilinx.DSP48E2}, this method allows to implement two multiplications $w_0 \cdot a_0 = r_0$ and $w_1 \cdot a_1 = r_1$ and a multiply-accumulate result $r_2 = w_0 \cdot a_1 + w_1 \cdot a_0$ . To achieve maximal utilization, the operands $w_0, w_1$ must have a bit width of 4 and the operands $a_0, a_1$ must have a bit width of 5, resulting in a bit width of $4+5=9$ for the results $r_0, r_1$.

Mert et al.~\cite{A.C.Mert.2018} propose a method to map two multiplications $c_0 \cdot a_0 = r_0$ and $c_1 \cdot a_0 = r_1$ to a single DSP. Note that $c_0$ and $c_1$ must be constants. Also note that the variable input $a_0$ is the same for both multiplications. This methods requires the constants to be decomposed into shift operations prior to the synthesis of the circuit. However, in many applications, the multiplication operands change during runtime, rendering the proposed method infeasible. 

Kalali and Van Leuken~\cite{KalaliVanLeuken.2021} extend the method of Mert et al.~\cite{A.C.Mert.2018} by using a table lookup technique for storing the decomposed constants. The lookup table allows the constants to be changed during runtime. In addition, an approximate computing technique is proposed to reduce the large overhead caused by this lookup table.

In the {Xilinx} white paper~\cite{Xilinx.INT8}, a method is proposed to implement two multiplications of the form $w_0 \cdot a_0 = r_0$ and $w_1 \cdot a_0 = r_1$ on a single DSP. In the remainder of this paper, this procedure will be referred to as INT8-packing. Note that the variable input $a_0$ is the same for both multiplications. In this method, all input operands have a bit width of 8, thus resulting in two 16 bit results. A very similar method is proposed by Lee et al.~\cite{DoubleMAC.2018}.

In the {Xilinx} white paper~\cite{Xilinx.INT4}, a method is proposed to implement four multiplications of the form $w_0 \cdot a_0 = r_0$, $w_1 \cdot a_0 = r_1$, $w_0 \cdot a_1 = r_2$, $w_1 \cdot a_1 = r_3$ on a single DSP. The input operands $w_0, w_1, a_0, a_1$ have a precision of 4 bit, thus resulting in four 8 bit results. In the remainder of this paper, this procedure will be referred to as INT4-packing.


\section{Preliminaries: INT4-Packing}\label{sec:preliminaries}
To generalize the ideas proposed by Xilinx Inc.~\cite{Xilinx.INT4}, we first give an overview over INT4-packing. Here, we also introduce our notation. INT4-packing basically computes the outer product of two vectors $\mathbf{a}$ and $\mathbf{w}$, with both vectors having two elements each:
\begin{equation}\label{eq:int4}
\mathbf{a} \cdot \mathbf{w}^\top = 
    \begin{bmatrix}
a_{0} \\
a_{1}
\end{bmatrix}
\cdot
\begin{bmatrix}
w_{0} \\
w_{1}
\end{bmatrix}^\top    
= 
\begin{bmatrix}
a_{0}w_{0} & a_{0}w_{1} \\
a_{1}w_{0} & a_{1}w_{1}
\end{bmatrix} 
\end{equation}

\begin{figure*}[!t]
  \centering
  \includegraphics[width=0.75\linewidth]{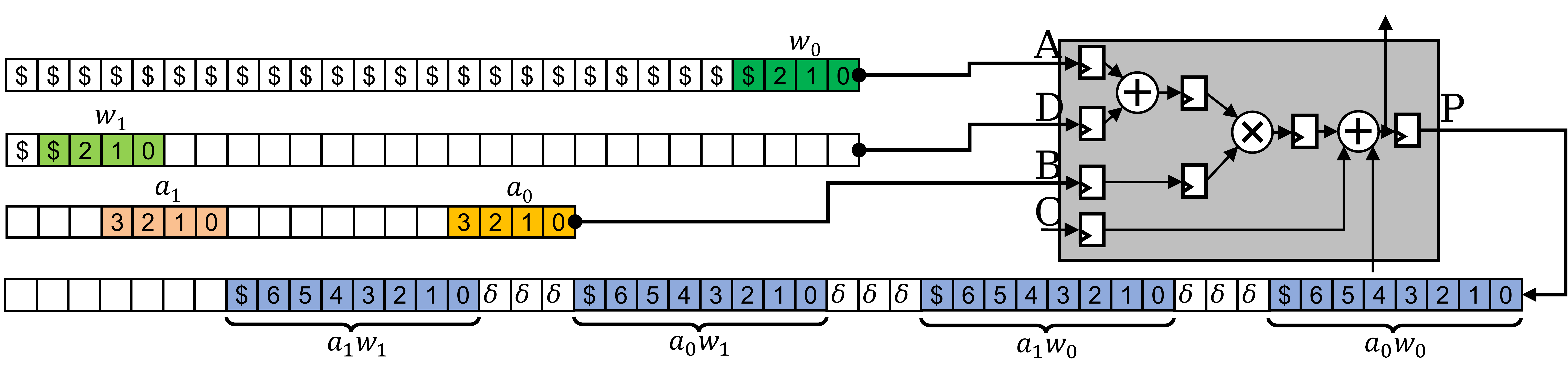}
  \caption{INT4-packing for performing four parallel multiplications on a single DSP. Here, \$ denotes the (extended) sign bits and $\delta$ denotes the padding bits. The offsets of the individual inputs determine the offsets of the individual results as described in Eqn.~\eqref{eq:int4_packing}.}
  \label{fig:int4_packing}
\end{figure*}

Computing the result of this outer product requires 4 multiplications in total. Using the INT4-approach, these multiplications can be packed on a single DSP if the entries of $\mathbf{a}$ are unsigned 4 bit integers and the entries of $\mathbf{w}$ are signed 4 bit integers. The strategy is to rearrange the individual inputs $a_{0}, a_{1}, w_{0}, w_{1}$ as described in the following equation:
\begin{multline}\label{eq:int4_packing}
(a_1 \cdot 2^{11} + a_0) \cdot (w_1 \cdot 2^{22} + w_0) \\ 
= a_1w_1 \cdot 2^{33} + a_0w_1 \cdot 2^{22} + a_1w_0 \cdot 2^{11} + a_0w_0
\end{multline}
Note that multiplications or divisions with $2^n$ can be implemented by fixed shift operations
that only require a rewiring of the individual bits. The computation in Eqn.~\eqref{eq:int4_packing} can be mapped to the DSP48E2 as follows:
The operand $a_0$ is mapped to the B-Port (see Fig.~\ref{fig:DSP48E2}) with an offset of 0. $a_1$ is also mapped to the B-Port but with an offset of 11. This is a hardware-efficient way of implementing $a_{1} \cdot 2^{11} + a_{0} \cdot 2^{0}$. Input $w_{0}$ is mapped to the preadder port A with an offset of zero. Since $w_{0}$ is signed, the sign bit has to be repeated for all Most Significant Bits (MSBs) to perform sign extension. Input $w_{1}$ cannot be mapped to the same port as $w_{0}$ because it is signed. Therefore, $w_{1}$ is mapped to the preadder port D with an offset of 22. The 4 results of the outer product can be extracted from the P-Port. E.g., the result $a_1w_1$ can be extracted from bit 33 to bit 40 from the P-Port. The individual results are separated by $\delta = 3$ padding bits  (see Fig.~\ref{fig:int4_packing}). This is important when multiple DSPs are chained together using the carry ports ($P_\text{in}, P_\text{cout}$) in order to accumulate their results.
Thus, with $\delta$ bits padding a maximum of $2^{\delta}$ results can be accumulated without error. When no results are accumulated no padding is needed.

\section{Generalization of INT4-Packing: INT-N}\label{sec:generalization}
The architecture-independent packing INT-N is used to generate a multiplication packing that
does not consider the constraints of the target DSP (i.e., limited bit widths and input ports). To generalize the INT4-approach, the multiplication packing technique has to be described in mathematical terms. For the generalization, the integer input vectors $\mathbf{a} = [a_0...a_n]$ and $\mathbf{w} = [w_0...w_m]$ can have \emph{any} number of entries. A packing configuration can then be described as follows. The offsets of the entries of $\mathbf{a}$ and $\mathbf{w}$ are stored in the sets $\mathbf{a}_{\mathrm{off}}$ and $\mathbf{w}_{\mathrm{off}}$ and the bit widths are stored in $\mathbf{a}_{\mathrm{wdth}}$ and $\mathbf{w}_{\mathrm{wdth}}$, respectively. The bit widths of the individual elements can be chosen arbitrarily, even allowing for cases where the each entry has a different bit width.
The result $\mathbf{r}$ contains the outer product, i.e., $\mathbf{r} = \mathbf{a} \cdot \mathbf{w}^\top$. Thus, the sets $\mathbf{r}_{\mathrm{off}}$ contains the offsets and $\mathbf{r}_{\mathrm{wdth}}$ contains the bit widths of the individual results. For example, INT4-packing has the following packing configuration (see Fig.~\ref{fig:int4_packing}):
Padding $\delta = 3$, $\mathbf{w}_{\mathrm{wdth}} = \mathbf{a}_{\mathrm{wdth}} = \{4,4\}$, $\mathbf{r}_{\mathrm{wdth}} = \{8,8,8,8\}$, $\mathbf{w}_{\mathrm{off}} = \{0,22\}$, $\mathbf{a}_{\mathrm{off}} = \{0,11\}$, $\mathbf{r}_{\mathrm{off}} = \{0,11,22,33\}$. 
The offsets of the inputs $\mathbf{a}_{\mathrm{off}}$ and $\mathbf{w}_{\mathrm{off}}$ determine the offsets of the results $\mathbf{r}_{\mathrm{off}}$ as described in Eqn.~\eqref{eq:feasible} (with $\vert\cdot\vert$ denoting the cardinality).
\begin{multline}\label{eq:feasible}
   \left(\sum_{i=0}^{\vert\mathbf{a}_{\mathrm{off}}\vert-1} a_i \cdot 2^{\mathbf{a}_{\mathrm{off},i}}\right)
   \cdot  
   \left(\sum_{j=0}^{\vert\mathbf{w}_{\mathrm{off}}\vert-1} w_j \cdot 2^{\mathbf{w}_{\mathrm{off},j}} \right)
   \\=
   \sum_{j=0}^{\vert\mathbf{w}_{\mathrm{off}}\vert-1}\sum_{i=0}^{\vert\mathbf{a}_{\mathrm{off}}\vert-1} a_iw_j \cdot 
   2^{\mathbf{r}_{\mathrm{off},j \cdot \vert\mathbf{a}_{\mathrm{off}}\vert + i}}
\end{multline}
Note that in the equation above, the offsets of the inputs $\mathbf{w}_{\mathrm{off}}, \mathbf{a}_{\mathrm{off}}$ determine the offsets of  the results $\mathbf{r}_{\mathrm{off}}$, i.e.:
$\mathbf{r}_{\mathrm{off},j \cdot \vert\mathbf{a}_{\mathrm{off}}\vert + i} = \mathbf{a}_{\mathrm{off},i} + \mathbf{w}_{\mathrm{off},j}$
This can be understood as a generalization of Eqn.~\eqref{eq:int4_packing}. Note that this generalized packing INT-N also applies to INT8 packing.

\section{Error analysis of multiplication packing}\label{sec:errors}
The generalized multiplication packing INT-N (and thus including the the INT4 and INT8 packing technique proposed by {Xilinx}) suffers from an error where some actual outputs $O_{\text {actual}}$ are smaller than the expected outputs $O_{\text {expect}}$. This error is bounded by $-1$, i.e., $O_{\text{actual}}$~=~$O_{\text{expect}}~-~1$. The error is introduced by the right-shifting operation that is implicitly performed when extracting the results from the bit string. The individual results
$a_iw_j$ are extracted from the result vector by right shifting. The reasoning behind this is that
\begin{equation}
x \gg n = \frac{x}{2^n}
\end{equation}
However, right shifting \emph{signed} integers implements division that is always rounding down i.e.:
\begin{equation}
x \gg n = \floor*{\frac{x}{2^n}} 
\end{equation}
This biases the output towards negative infinity.
For INT4-packing, this leads to an error rate of around 37\% over all possible input combinations.

We propose two error correction schemes: one method for full error correction requiring extra hardware (LUTs, FFs) and one approximate method that reduces the error probability from 37\% to 3\% and requires no additional hardware. 
\subsection{Full Error Correction}
Since the errors are related to rounding, they can be fixed by applying the correct rounding.
The correct rounding should implement rounding that does not introduce a bias. For this,
rounding to the nearest integer is used. This rounding scheme can be implemented using
the Round-Half-Up function:
\begin{equation}
\mathrm{round\_half\_up} = \lfloor x+0.5 \rfloor
\end{equation}
For example, 3.1 gets rounded to 3, and 3.6 gets rounded to 4. The advantage of this
function is that it can be implemented easily in hardware by checking a single bit. To
implement the Round-Half-Up function, the cases where the result needs to be rounded up
have to be determined, because rounding down is done by default. For this, the results can
be interpreted as fixed-point numbers:
\begin{itemize}
    \item If the first bit behind the decimal point is 1, then round up by adding 1 to the result.
    \item Else, round down by adding a 0 (i.e. do nothing).
\end{itemize}

\begin{figure*}[!t]
  \centering
  \includegraphics[width=0.75\linewidth]{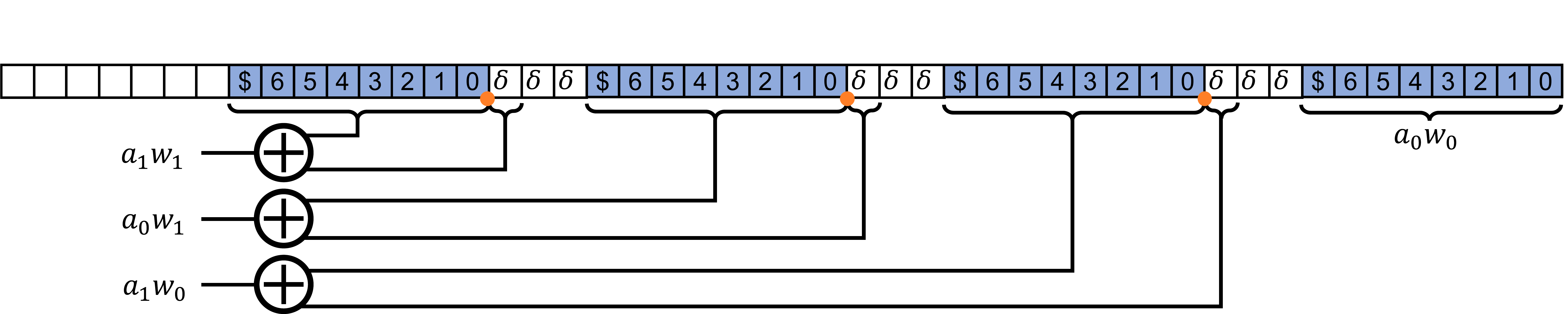}
  \caption{Hardware circuit for implementing the round half up function using additional hardware. The orange dots denote the imaginary decimal points of the results that have to be extracted.}
  \label{fig:full_correct}
\end{figure*}
Full error correction requires additional hardware resources for implementing the required adders (see Fig.~\ref{fig:full_correct}).

\subsection{Approximate Correction}\label{sec:approx_correct}
\begin{figure*}[!t]
  \centering
  \includegraphics[width=0.75\linewidth]{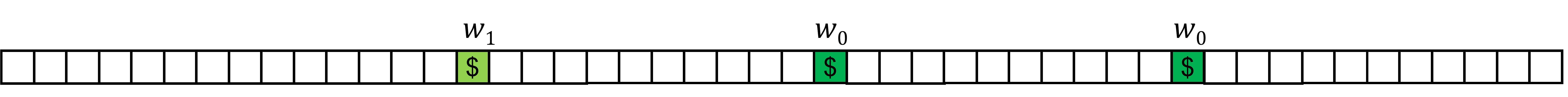}
  \caption{48-Bit correction term that is fed into the DSP's accumulator port C for the approximate error correction technique. Shown here for INT4 packing.}
  \label{fig:approx_correct}
\end{figure*}

The idea is to anticipate if the first bit after the decimal point is 1 (thus requiring rounding up) and then adding a 1 to the result \emph{before} extracting to result. The proposed solution works as follows. The assumption is that  $\mathbf{a}$ contains unsigned entries and $\mathbf{w}$ contains signed entries. The first bit after the decimal point of the $n$'th result located at $\mathbf{r}_{\mathrm{off},n}$ is 1 if the sign of the result located at $\mathbf{r}_{\mathrm{off},n-1}$ is negative (due to the sign bits). The ideas is then that the sign of a result located at $\mathbf{r}_{\mathrm{off},n-1}$ can be anticipated by checking the sign of the operands.
Thus, adding the sign bit of the operands that generated the result located at $\mathbf{r}_{\mathrm{off},n-1}$ resolves some of the errors. This addition can be performed using the DSP's accumulator that is accessed through the C-Port (see Fig.~\ref{fig:approx_correct}). In some rare cases, this does not resolve the errors, e.g. when one operand is zero.

\section{Overpacking}\label{sec:Overpacking}
In the domain of digital signal processing, many applications do not require computations
to be exact. Reducing computational accuracy often allows for improvements on
energy efficiency, resource utilization or execution speed~\cite{Xu.2016}. If approximation
is acceptable and to what degree depends on the application. Typical applications are in
the domain of image processing and machine learning, since the underlying algorithms in
these domains are often of an approximate nature themselves.
The general idea for approximate multiplication on DSPs is to reduce the amount of
padding bits $\delta$ below the minimum.
If $\delta$ is reduced below the minimum, some bits of the individual results are being merged together and errors occur. The benefit of this technique is that more multiplications with a higher bit precision can be fitted on a single DSP block.
The minimum required amount of $\delta$-bits is determined by how many results are supposed to be accumulated. For easier interpretation of the results, Overpacking experiments have been performed with no accumulation of results. Therefore, the minimum padding is $\delta = 0$. Thus, Overpacking can be introduced by setting $\delta = -1, -2,$ etc.

\subsection{Errors introduced by Overpacking}
\begin{figure}[!t]
    \centering
    \begin{subfigure}{.20\textwidth}
      \centering
      \includegraphics[width=1\textwidth]{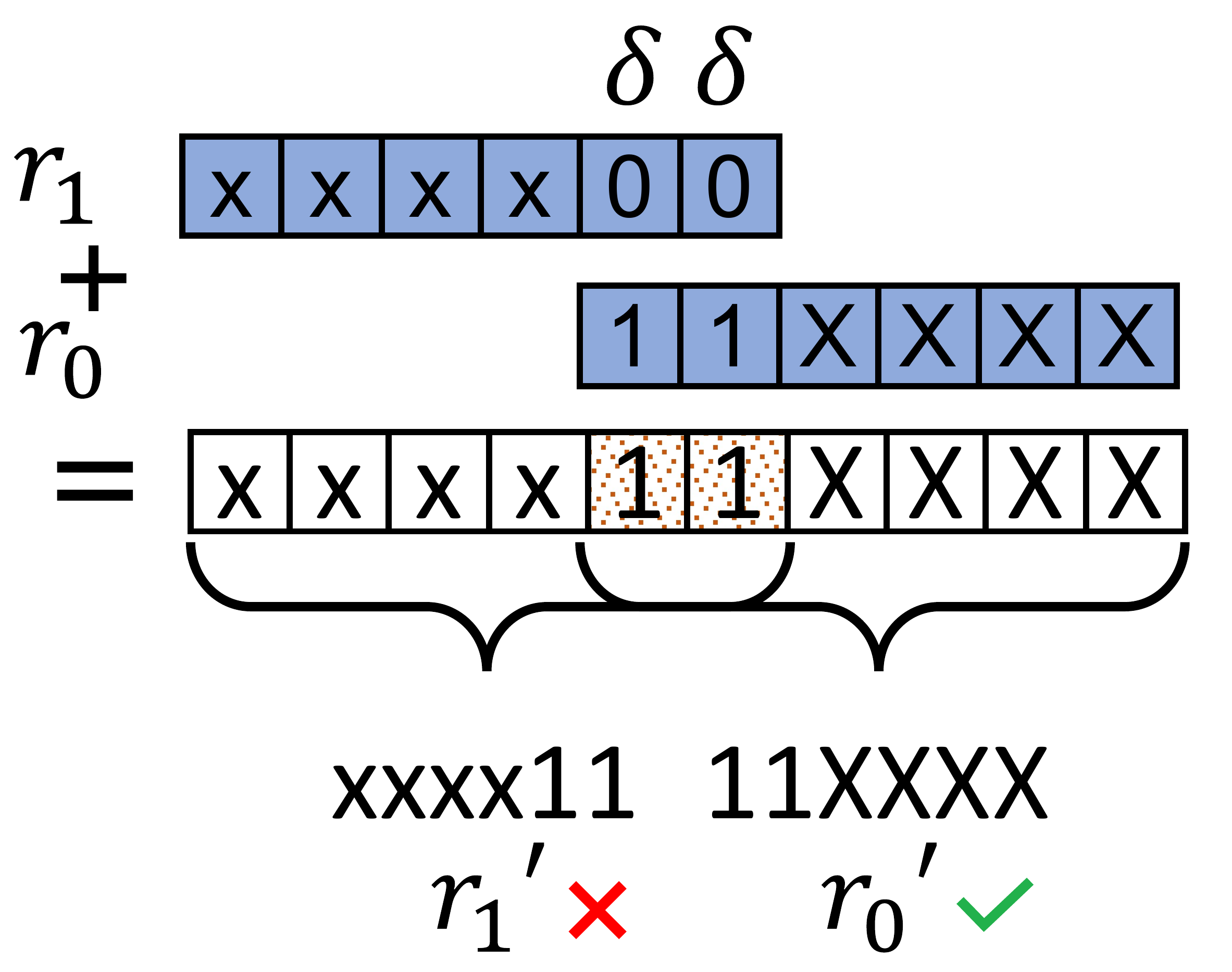}
      \caption{LSB contamination}
      \label{fig:lsb_contamination}
    \end{subfigure}%
    \hfill
    \begin{subfigure}{.20\textwidth}
      \centering
      \includegraphics[width=1\textwidth]{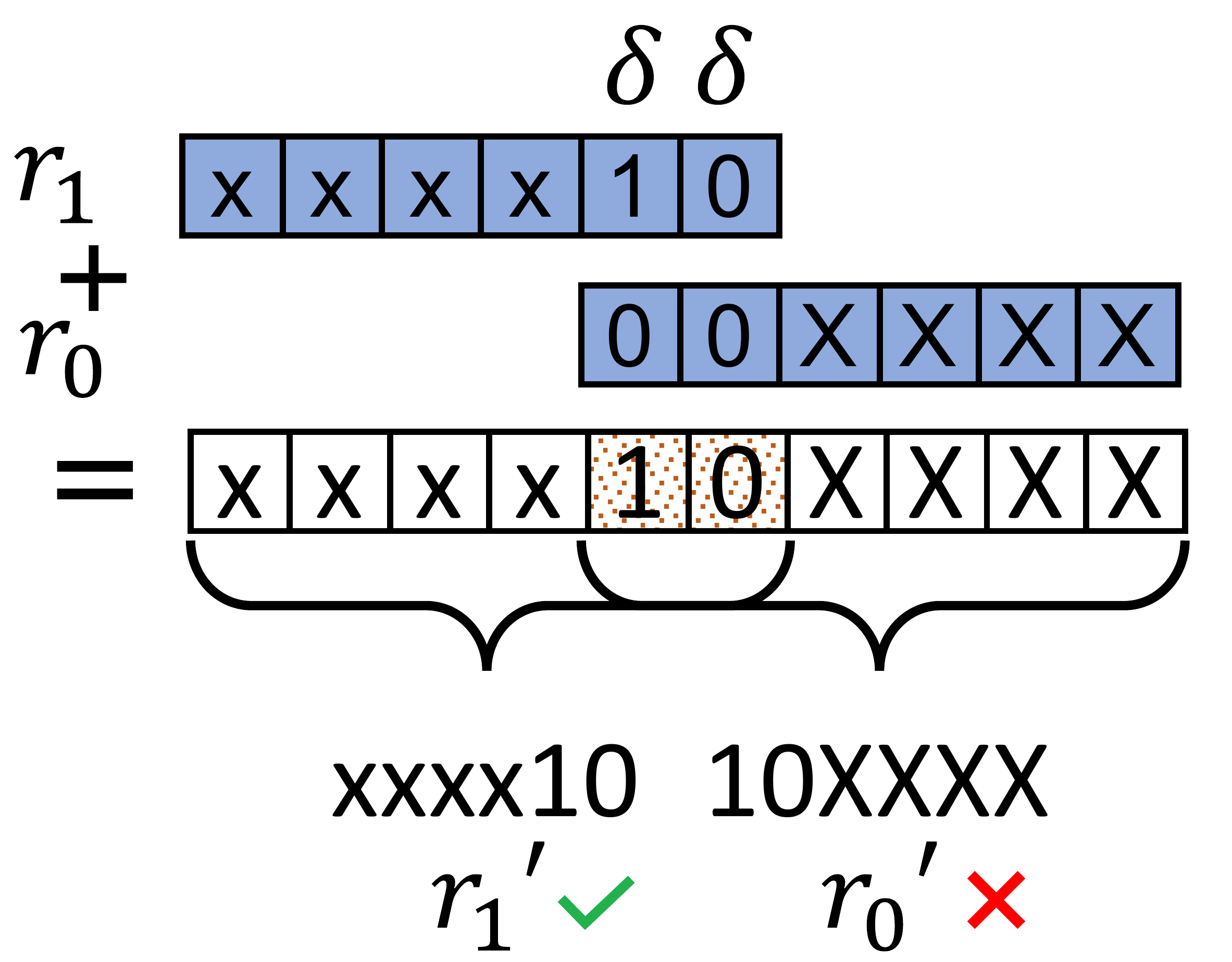}
      \caption{MSB contamination}
      \label{fig:msb_contamination}
    \end{subfigure}%
    \caption{Errors introduced by Overpacking. The result $r_1$ and $r_0$ overlap because of the negative padding $\delta=-2$. $r_1'$ and $r_0'$ are the corrupted results extracted from the output vector. X and x denote don't care bits of $r_1$ and $r_0$, respectively. }
    \label{fig:Overpacking_error}
\end{figure}

For Overpacking, the individual results are separated by less $\delta$-bits than required. Thus, the individual results $a_iw_j$ overlap. The mathematical operation that fuses the overlapping
results to the final output vector P is an addition (according to Eqn.~\eqref{eq:feasible}). This has the effect that the LSBs of a results $r_1$ located left of a result $r_0$ affect the MSBs of $r_0$. For example, Fig.~\ref{fig:lsb_contamination} shows how non-zero MSBs of $r_0$ contaminate the LSBs of $r_1$. Fig.~\ref{fig:msb_contamination} shows how non-zero LSBs of $r_1$ contaminate the MSBs of $r_0$. An approximate computing technique where the MSBs of a result are contaminated leads to very high errors  and is therefore not desirable.
We present a technique to correct these errors with very little hardware overhead.

\subsection{MR-Overpacking}
Most Significant Bit Restoring (MR) Overpacking is an improvement to the naive Overpacking technique introduced above. The idea of MR-Overpacking is to let the MSBs get contaminated (Fig.~\ref{fig:msb_contamination}), but to restore the MSBs after extraction. The reasoning behind this is: erroneous MSBs lead to a high error, erroneous LSBs are not having a large impact on the result.
To restore the MSBs, the \emph{inverse} process of the MSB contamination must be performed. 
As illustrated in Fig.~\ref{fig:msb_contamination}, $\delta$ MSBs of a result $r_n$ that is located at $\mathbf{r}_{\mathrm{off},n}$ are contaminated by adding $\delta$ LSBs of another result $r_{n+1}$ that is located at $\mathbf{r}_{\mathrm{off},n+1}$. Thus, the MSB contamination is an \emph{addition} which can be inverted using a \emph{substraction}.
MR-Overpacking restores the $\delta$ MSBs of a result $\mathbf{r}_{\mathrm{off},n}$ by \emph{substracting} the $\delta$ LSBs of $\mathbf{r}_{\mathrm{off},n+1}$. 
To perform the restoration, the contaminating LSBs have to be known. Therefore, the contaminating LSBs must be calculated. The first LSB (Eqn.~\eqref{eq:lsb0}) and the second LSB (Eqn.~\eqref{eq:lsb1}) of a  result can be calculated with a little  hardware overhead. Eqns.~\eqref{eq:lsb0}, \eqref{eq:lsb1} are simply the rules for binary multiplication.

\begin{equation}\label{eq:lsb0}
  a_iw_j[0] = a_i[0] \wedge w_j[0]
\end{equation}
\begin{equation}\label{eq:lsb1}
  a_iw_j[1] = (a_i[0] \wedge w_j[1]) \oplus (a_i[1] \wedge w_j[0])
\end{equation}

\begin{figure*}[!t]
  \centering
  \includegraphics[width=0.75\linewidth]{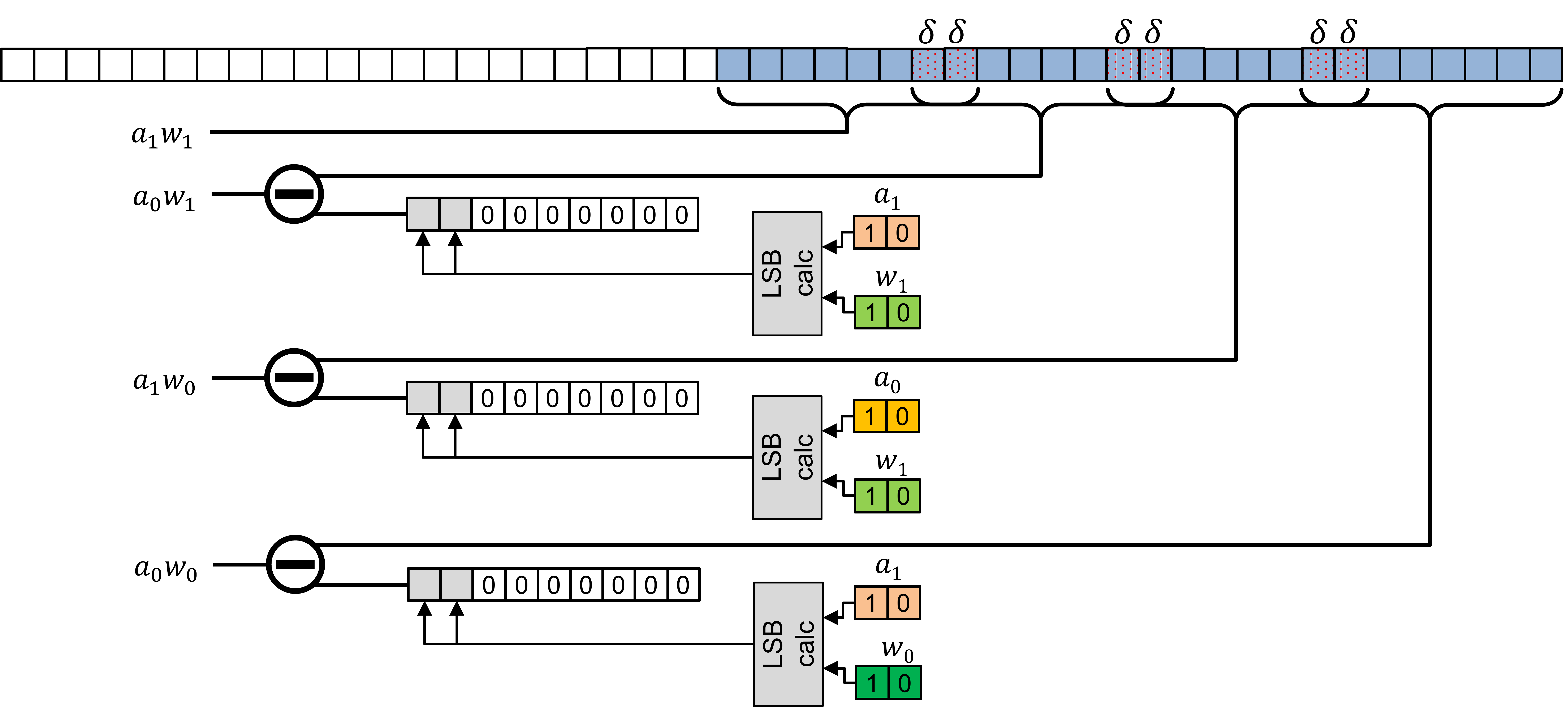}
  \caption{Hardware circuit for implementing MR-Overpacking with $\delta = -2$. The boxes ``LSB calc'' implement the Eqns.~\eqref{eq:lsb0},~\eqref{eq:lsb1} for the construction of the correction terms that are then subtracted from the extracted results. To obtain this packing, the following setup is used: $\mathbf{w}_{\mathrm{wdth}} = \mathbf{a}_{\mathrm{wdth}} = \{4,4\}$, $\mathbf{r}_{\mathrm{wdth}} = \{8,8,8,8\}$, $\mathbf{w}_{\mathrm{off}} = \{0,12\}$, $\mathbf{a}_{\mathrm{off}} = \{0,6\}$, $\mathbf{r}_{\mathrm{off}} = \{0,6,12,18\}$.}
  \label{fig:mr_Overpacking}
\end{figure*}

Consider the following example where $\delta=-2$, $a_0 = 1010_2 = 10_{10}$, $a_1 = 0011_2 = 3_{10}$, $w_0 = 1001_2 = -7_{10}$, $w_1 = 1100_2 = -4_{10}$, with $w_0, w_1$ being signed integers in 2's complement notation. For Overpacking, set $\delta~=~-2$. The expected result for $a_0w_0$ would be $1011~1010_2~=-70_{10}$. However, due to Overpacking, $a_0w_0'~=~0111~1010_2=122_{10}$. The two LSBs of $a_1w_0$ that corrupt $a_0w_0$ are calculated using Eqn.~\eqref{eq:lsb0},~\eqref{eq:lsb1}: $a_1w_0[0] = a_1w_0[1] = 1$. The correction is performed by subtracting the corrupting LSBs : $a_0w_0 = a_0w_0' - 1100~0000_2$. Fig.~\ref{fig:mr_Overpacking} shows how this scheme can be implemented with extra hardware (LUTs and FFs). Note that this scheme must be adapted according to $\delta$. For example, if $\delta=-1$, only the implementation of Eqn.~\eqref{eq:lsb0} is required. When $\delta$ is increased further, the respective LSBs must be calculated according to the rules of binary multiplication. For example, setting $\delta~=~-4$ requires the calculation of the first four LSBs.  Note  that the hardware cost for implementing the correction logic increases exponentially with the number of LSBs.

The catch of the proposed scheme is as follows. Multiplication is very expensive in terms of hardware resources with two exceptions: the first and the second LSBs are very cheap to implement (see Eqns.~\eqref{eq:lsb0},~\eqref{eq:lsb1}). This means, that the expensive multiplication is performed in the optimized DSPs, the correction of corrupted MSBs is cheap to implement with additional hardware. The corruption of the LSBs only leads to small errors, as will be discussed in Section~\ref{sec:experiments}. 

\section{Addition Packing}\label{sec:addition_packing}
\begin{figure}[!t]
  \centering
  \includegraphics[width=0.75\linewidth]{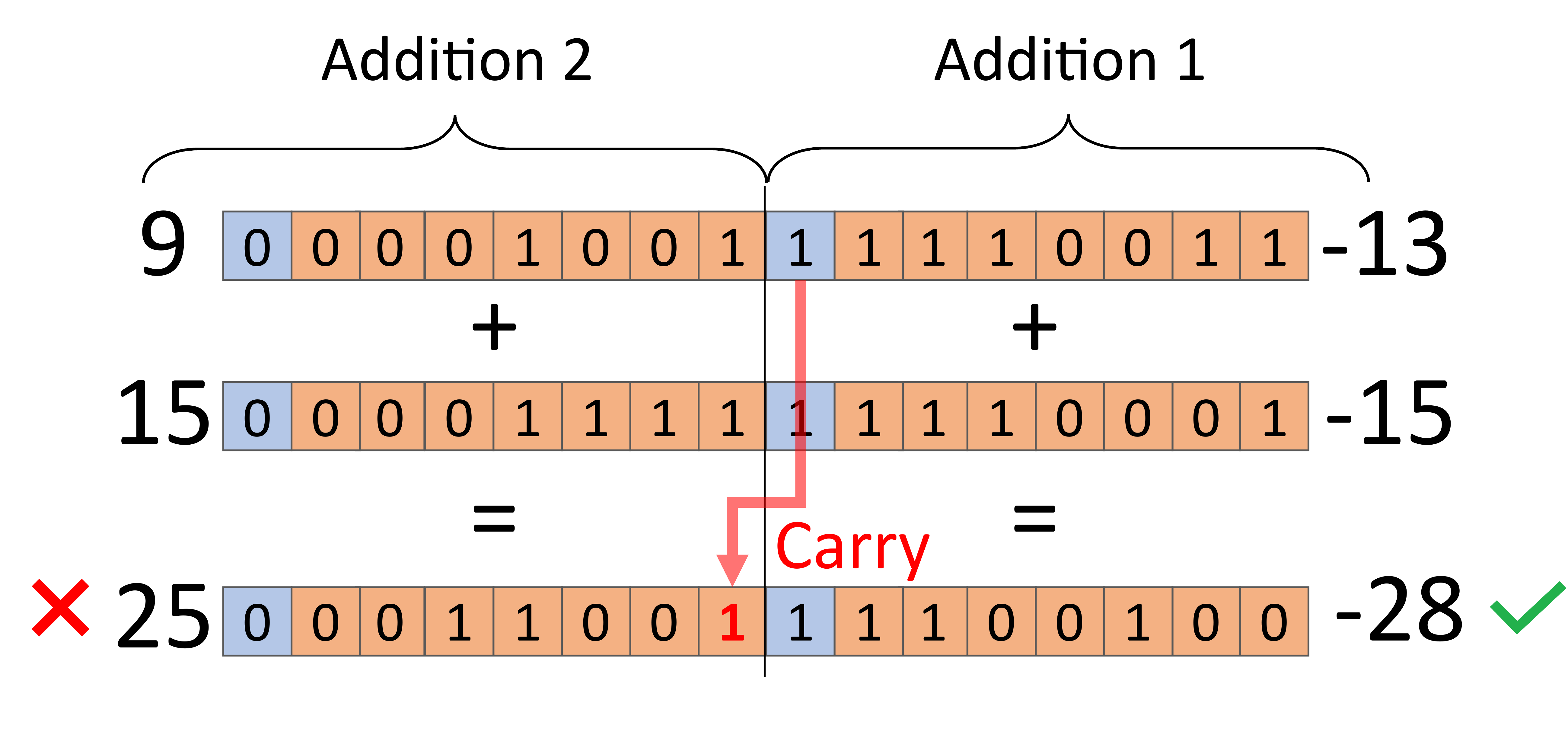}
  \caption{Strategy for packing two small bit with additions to a large one. Shown is the origin of the error where the carry signal of addition 1 propagates to the least significant bit of addition 2.}
  \label{fig:addition_packing}
\end{figure}

\begin{figure}[!t]
  \centering
  \includegraphics[width=0.75\linewidth]{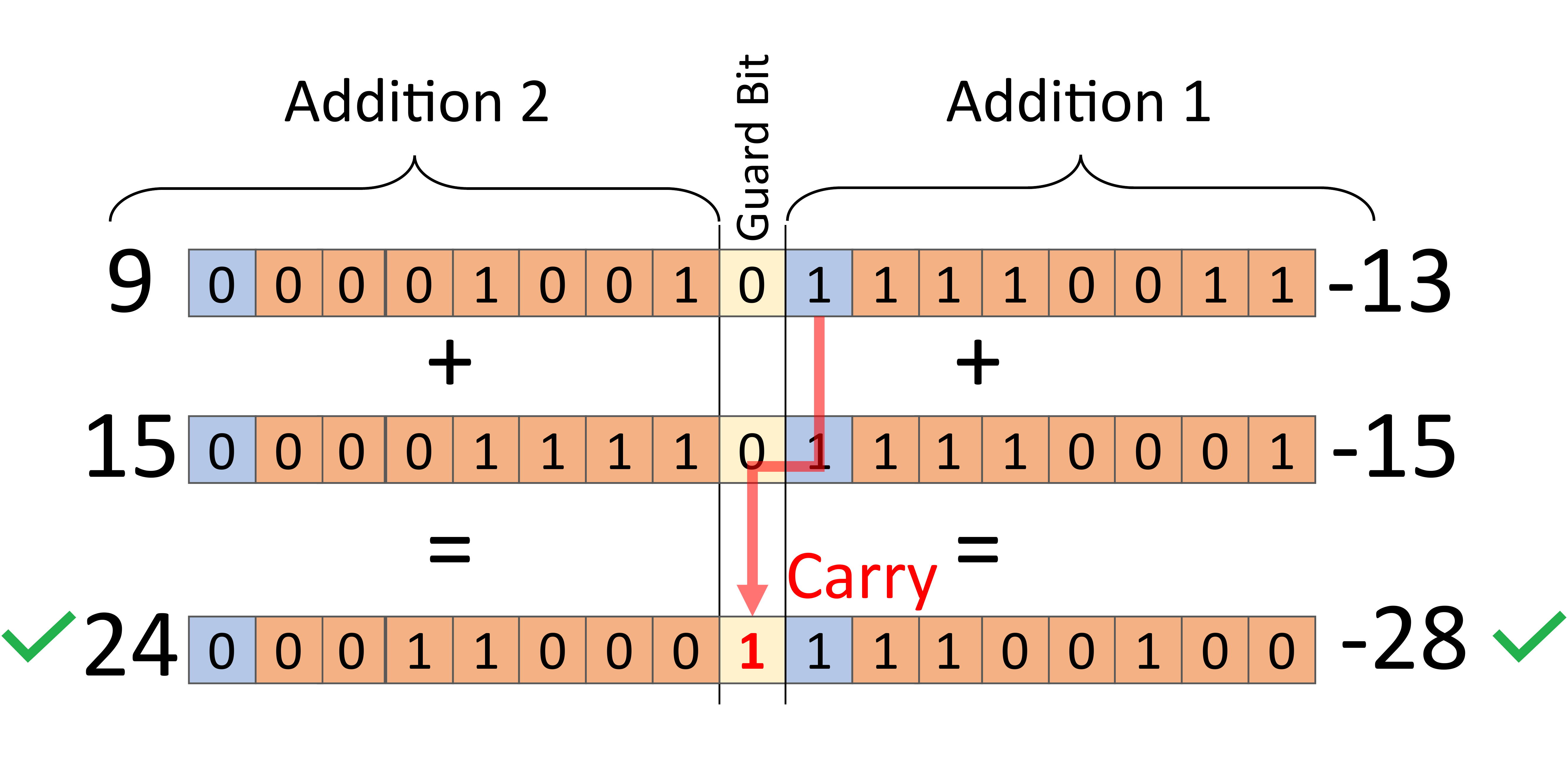}
  \caption{Addition packing with no errors using a guard bit between the individual additions. The guard bit ``catches'' the carry signal and prevents it from propagating to addition 2.}
  \label{fig:addition_packing_guard_bit}
\end{figure}

In Spiking Neural Networks (SNNs), the main computational operation is an addition~\cite{Bouvier.2019, Rueckauer.2017}. In contrast, standard Neural Networks (NNs) are based on Multiply-Accumulate operations (MAC)~\cite{Sze.2017b}. This means that SNN-accelerators require the implementation of thousands of adders, placing a high load on the FPGA's LUT and FF resources. In the following, we propose a method to pack multiple small-bit additions into the large 48-bit adders of the DSP48. Consider the example in Fig.~\ref{fig:addition_packing} where two 8 bit additions are mapped to a single 16 bit addition. One 8 bit addition is mapped to the lower 8 bits and the other 8 bit addition is mapped to the upper 8 bits. The addition in the upper bits is still connected to the addition in the lower bits by the carry chain. Thus, if the addition in the lower 8 bits causes a carry, this carry is propagated to the addition in the upper 8 bits. The propagated carry causes an error in the least significant bit of the upper addition. This has the following implications: a) the addition located in the lower bits never has errors. (b) If a carry occurs from one addition to another, only the least significant bit of a result is corrupted. Therefore, the worst case absolute error is bounded to 1. To completely prevent errors, an extra guard bit with the value of 0 can be added between all additions. This guard bit is used to prevent that a carry signal can propagate from one addition to another (as illustrated in Fig.~\ref{fig:addition_packing_guard_bit}). This guard bit is additional overhead as it cannot be used for addition. This scheme can be generalized to different bit widths and more additions. The DSP48 features a 48-bit adder. For example, five 9 bit adders can be packed into a single DSP leaving room for three guard bits. Therefore, only a single adder is approximating in the sense that some results may have errors. To maximize the utilization, two 9-bit adders and three 10-bit adders can be mapped to a single DSP, leaving no space for guard bits.

\section{Experiments and Results}\label{sec:experiments}
To analyze the techniques and their errors described in the previous chapters, the following error metrics are used to compare an an expected output $O_{\text {expect}}$ to an actual output $O_{\text {actual}}$~\cite{V.Mrazek.2017}.
All $N$ possible input combinations were tested.

\begin{equation}
\operatorname{EP}=\frac{\sum_{\forall n: O_{\text {expect }}^{(n)} \neq O_{\text {actual }}^{(n)}}^{N} 1}{N} \cdot 100 \%
\end{equation}

\begin{equation}
\operatorname{MAE}=\frac{1}{N}\cdot \sum_{n=1}^N\left|O_{\mathrm{actual }}^{(n)}-O_{\mathrm{expect }}^{(n)}\right|
\end{equation}\label{eq:me}

\begin{equation}
\mathrm{WCE}=\max _{\forall n}\left|O_{\mathrm {actual }}^{(n)}-O_{\mathrm {expect }}^{(n)}\right|
\end{equation}\label{eq:wce}

\begin{figure}[!t]
\centering
\begin{tikzpicture}
\begin{axis}[
xbar,
width=7cm, height=4cm, enlarge x limits=0.5, /pgf/bar width=5pt,
xlabel={Packing Density $\rho$},
symbolic y coords={{INT4},INT-8,Huang et al.,INT-N,Overpacking},
ytick=data,
nodes near coords, nodes near coords align={horizontal},
]
\addplot coordinates {(0.67,{INT4}) (0.67,INT-8) (0.56,Huang et al.) (0.875,INT-N)(1.13,Overpacking)};
\end{axis}
\end{tikzpicture}
\caption{Multiplication packing density for different techniques.}
\label{fig:density}
\end{figure}
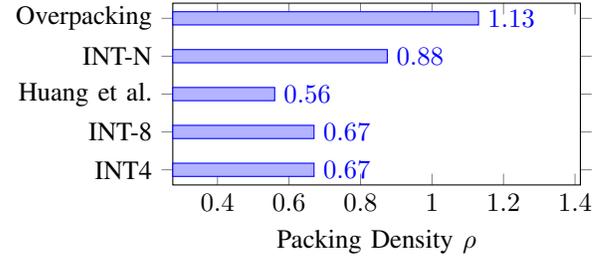

Note that the error statistics listed above are calculated for each result $a_iw_j$ individually. The result calculated over all individual results is indicated by a bar accent (e.g., $\overline{\operatorname{MAE}}$). 
To evaluate the hardware cost of the individual circuit implementations, each circuit was implemented on the the Xilinx Zynq UltraScale+ MPSoC (XCZU7EV-2FFVC1156). For best comparability with the related INT4-approach by Xilinx, all tests in Table~\ref{tab:packing_results} and~\ref{tab:MR_packing_results} have been performed with 4-bit operands and four multiplications. Note that the approaches introduced in this paper can be generalized to arbitrary bit widths and number of multiplications, restricted only by the DSPs architecture. Table~\ref{tab:addition_packing} shows the result of a single 9-bit adder that is implemented along with four other 9-bit adders in a single DSP using the addition packing scheme without guard bits. 
To measure the effectiveness of the different packing schemes, we introduce the packing density $\rho = b_\text{used}/b_\text{total}$, where $b_\text{total}$ is the number of output bits (e.g., 48 bits for the DSP48) and $b_\text{used}$ is the number of bits that are occupied by the multiplication results. 
Fig.~\ref{fig:density} compares the multiplication packing density for the different approaches. To evaluate INT-N, the following packing configuration is used: Padding $\delta = 0$, $\mathbf{w}_{\mathrm{wdth}} = \{3,3\}$ $\mathbf{a}_{\mathrm{wdth}} =\{4,4,4\} $, $\mathbf{r}_{\mathrm{wdth}} = \{7, 7, 7, 7, 7, 7\}$, $\mathbf{w}_{\mathrm{off}} = \{0, 21\}$, $\mathbf{a}_{\mathrm{off}} = \{0, 7, 14\}$, $\mathbf{r}_{\mathrm{off}} = \{0, 7, 14, 21, 28, 35\}$. 
The packing configuration for Overpacking is: $\delta = -2$, $\mathbf{w}_{\mathrm{wdth}} = \{5,5\}$ $\mathbf{a}_{\mathrm{wdth}} =\{4,4,4\} $, $\mathbf{r}_{\mathrm{wdth}} = \{9, 9, 9, 9, 9, 9\}$, $\mathbf{w}_{\mathrm{off}} = \{0, 21\}$, $\mathbf{a}_{\mathrm{off}} = \{0, 7, 14\}$, $\mathbf{r}_{\mathrm{off}} = \{0, 7, 14, 21, 28, 35\}$. Note that many more configurations for INT-N and Overpacking are possible (Eqn ~\eqref{eq:feasible}), that can be adapted according to the application at hand.

\begin{table}[!t]
  \caption{Results of the various presented multiplication packing approaches, compared to the approach proposed by Xilinx~\cite{Xilinx.INT4}. All operands are 4-bit with four multiplications on a single DSP.}
  \label{tab:packing_results}
  \centering
  \begin{tabular}{lccccc}
    \toprule
        Approach                    & $\overline{\operatorname{MAE}}$             & $\overline{\operatorname{EP}}$      & $\overline{\operatorname{WCE}}$     & LUTs       & FFs  \\
    \midrule    
    Xilinx INT4~\cite{Xilinx.INT4} & 0.37            & 37.35\% &  1      & 0          & 0   \\
    INT4 Full Correction           & 0.00            & 0.00\%  &  0      & 27         & 32  \\
    INT4 Approx. Correction        & 0.02            & 3.13\%  &  1      & 0          & 0   \\
    Overpacking $\delta=-1$         & 24.27           & 49.85\% &  129    & 0          & 0   \\
    Overpacking $\delta=-2$         & 37.95           & 58.64\% &  194    & 0          & 0   \\
    Overpacking $\delta=-3$         & 45.53           & 78.26\% &  228    & 0          & 0   \\
    MR-Overpacking $\delta=-1$      & 0.37            & 37.35\% &  1      & 4          & 6   \\
    MR-Overpacking $\delta=-2$      & 0.47            & 41.48\% &  2      & 6          & 20   \\
    MR-Overpacking $\delta=-3$      & 0.78            & 49.95\% &  4      & 17         & 30   \\
  \bottomrule
\end{tabular}
\end{table}

\begin{table}[!t]
  \caption{Detailed error statistics of the individual results $a_iw_j$ for INT4 Packing and MR-Overpacking.}
  \label{tab:MR_packing_results}
  \centering
  \begin{tabular}{lcccccc}
    \toprule
                                    & \multicolumn{3}{c}{INT4 Packing} & \multicolumn{3}{c}{MR-Overpacking $\delta = -2$} \\ 
    \cmidrule(r){2-4} \cmidrule(l){5-7}
    Result                          & MAE       & EP               & WCE & MAE       & EP               & WCE            \\
    \midrule                                                                                                    
    $a_0w_0$                        & 0.00      & 0.00\%           & 0   & 0.00      & 0.00\%           & 0      \\ 
    $a_1w_0$                        & 0.47      & 46.87\%          & 1   & 0.60      & 52.34\%          & 2        \\ 
    $a_0w_1$                        & 0.50      & 49.80\%          & 1   & 0.64      & 55.41\%          & 2      \\ 
    $a_1w_1$                        & 0.53      & 52.73\%          & 1   & 0.66      & 58.20\%          & 2      \\
    for all $a_iw_j$                & 0.37      & 37.35\%          & 1   & 0.47      & 41.48\%          & 2      \\
  \bottomrule
\end{tabular}
\end{table}

\begin{table}[!t]
  \caption{Results of the addition packing scheme.}
  \label{tab:addition_packing}
  \centering
  \begin{tabular}{lccccc}
    \toprule
        Approach                    & MAE             & EP      & WCE     & LUTs       & FFs  \\
    \midrule    
    Addition Packing                & 0.51            & 51.83\% &  1      & 0          & 0   \\
  \bottomrule
\end{tabular}
\end{table}

\section{Conclusion}\label{sec:conclusions}
DSP-Packing offers a way to effectively use DSPs for small width multiplications. 
The INT4-approach as proposed by Xilinx~\cite{Xilinx.INT4} for multiplication packing has multiple limitations. First, the bit widths of the individual multiplications and the amount of operands is fixed. 
The padding is $\delta$ is also fixed to 3 and it introduces a rounding-related error. 
We showed that this technique can be generalized to arbitrary packings.  
While the rounding error introduced by INT4-packing is relatively small (MAE = 0.37), it adds a small bias towards negative infinity to the results which might be an issue for some applications. 
The full error correction requires the implementation of additional adders, introducing an overhead as additional resources are required. Approximate error correction uses the internal accumulator of the DSP48E2, requiring no external hardware.  The error that is still present is very small ($\overline{\operatorname{MAE}} = 0.02$) making approximate error correction an ideal solution. 
Overpacking can be used to improve the utilization of DSP blocks significantly. 
Overpacking itself introduces a very large error since the MSBs of the extracted results are corrupted. 
MR-Overpacking reduces the error significantly with a very small hardware overhead.
This makes MR-Overpacking an interesting solution, for example for Convolutional Neural Networks (CNNs), because they are inherently resilient to quantization and approximations. 
For example, MR-Overpacking can be used to map 6 individual 4-bit multiplications on a single DSP48E2, allowing 50\% more 4-bit multiplications compared to INT4-packing while achieving the same $\overline{\operatorname{MAE}} = 0.37$.
MR-Overpacking can also be used to increase the precision of multiplications. 
For example, setting $\delta=-2$ allows 4 individual 6-bit multiplications on a single DSP, resulting in an increase in bit precision of 50\% compared to INT4-packing.
In the future, we plan to explore methods to dynamically change the DSP packing during runtime according to the requirements of the computational task at hand.

\bibliographystyle{IEEEtran}
\bibliography{main.bib}

\vfill

\end{document}